\documentstyle{article}
\def\sec#1\par{\let\Large=\relax\section{#1}}
\def\bi{\begin{list}{$\bullet$}{\topsep0pt \parsep=\parskip \itemsep=0pt}
\def\y{\item}}
\def\ei{\end{list}}
\def\label#1{{\bf [#1]}}
\def\ref#1#2{#2}
\def\[{$$}
\def\]{$$}
\def\d{\hbox{d}}
\def\mede#1{\left\langle#1\right\rangle}
\begin{document}
\advance\baselineskip by4pt
\vskip2mm
\centerline{\uppercase{\bf
Life-times of simulated traffic jams
}}
\vskip2mm
\centerline{Kai Nagel}
\centerline{Zentrum f\"ur Paralleles Rechnen,
c/o Math.~Inst.,}
\centerline{Universit\"at, 50923~K\"oln, Germany}
\centerline{{\tt kai@mi.uni-koeln.de}}
\centerline{\today}
\vskip2mm
\noindent{\leftskip1cm{\bf Abstract:}
We study a  model for   freeway traffic  which includes  strong  noise
taking  into account the  fluctuations of individual driving behavior.
The model shows emergent  traffic jams with a  self-similar appearance
near the throughput maximum of the traffic.  The lifetime distribution
of  these jams shows  a short scaling  regime, which gets considerably
longer if  one reduces the  fluctuations when driving at maximum speed
but  leaves  the   fluctuations   for slowing   down  or  accelerating
unchanged.
The outflow from a traffic jam
self-organizes into this state of maximum throughput.

}

\sec Introduction

Freeway  traffic consists of  (at least) two  different regimes, which
are (i)~dense traffic, where the individual velocity of each driver is
strongly  influenced by the presence  of other vehicles, and (ii)~free
traffic, where the presence of other vehicles  has no influence on the
speed.  It has  been    argued from  mathematical models of    freeway
traffic~\cite{LW55} that the change-over  from one regime to the other
might  be similar  to  a transition from   laminar to  turbulent fluid
flow~\cite{Kue91,PH88}.  In these mathematical models, traffic flow is
treated similar to a fluid flowing down  a narrow inclined channel.  A
typical phenomenon connected with this  change-over are shock-waves of
vehicles.  But this behavior  is not restricted  to fluids or traffic:
It  is common in many transportation  mechanisms, as, e.g., in granular
materials~\cite{Mehta,Poeschl,Lee}.

As thorough ``laboratory'' experiments with traffic are difficult to
undertake, especially with the large number of cars which would be
necessary for a meaningful treatment as a many-body problem, it makes
sense to work with simulation models which, in addition, do not rely on
the fluid assumption.  We use a seven state model on a
one-dimensional array similar to a cellular automaton~\cite{Wolfram}
which allows to obtain meaningful
results already on a workstation.  In addition, we use
high performance
computers in order to obtain data of a higher quality.

Standard models for  microscopic traffic  simulations~\cite{MJH90} are
not only computationally slower, but, as a  result of their attempt to
contain most aspects of real world traffic, are much more complicated.
As  we already have shown~\cite{NaS92}  that  even a most simple model
for single lane traffic captures many aspects  of reality, we continue
in this paper our investigation of  this model's behavior.  We believe
that our results   may be used   as a tool  to  better understand  the
corresponding structures  in    real world traffic.    This   paper is
complemented  by several   others  which  give results   on  numerical
performance       on   parallel  computers~\cite{NaS93},    analytical
treatment~\cite{ScS93,INSS93},  and   deterministic versions   without
noise~\cite{NaH93}. Further work, especially on multi-lane traffic, is
in  progress~\cite{NaR93}.  Similar approaches   are also  useful  for
understanding   the  principal properties     of   traffic  in    road
networks~\cite{Biham,spanish}.

The structure of this paper is as
follows: We start (section~2) with a
recapitulation of the essential features of our freeway traffic model.
Section~3 describes observations from density plots.  In
particular, these plots
show some self-similarity of the waves when the system operates at
the throughput maximum.
In section~4 we show numerically that the outflow of a jam
evolves automatically towards this state of maximum vehicle
throughput.
The succeeding two sections describe our attempts to quantify
this observation by measuring the lifetime distribution of the
jams: Section~5 contains the description of the algorithm,
and section~6 gives results.
In particular we find that a short scaling regime appears when
approaching the regime of maximum vehicle throughput.
A modification of the model leading
to a longer scaling region is discussed in section~7,
section~8 discusses possible implications for real world
traffic, and section~9 summarizes the results.

\sec Recapitulation of the single lane freeway traffic model

Our     freeway traffic   model     has   been  described in    detail
in~\cite{NaS92}.  Therefore, we only  want to give  a short account of
the essentials.  The single lane version of the model  is defined on a
one-dimensional   array of  length $L$,  representing  a (single-lane)
freeway.  Each site of the  array can only  be in one of the following
seven states: It may be occupied by one car having an integer velocity
between  zero and five, or  it may be empty.   This integer number for
the velocity  is the  number of sites  each vehicle  moves during  one
iteration; before   the movement, rules  for velocity  adaption ensure
``crash-free''  traffic.  The choice   of five as  maximum velocity is
somewhat   arbitrary, but it can  be  justified  by comparison between
model and   real   world measurements,  combined   with the    aim for
simplicity of the model.  In any case, any value $v_{max} \ge 2$ seems
to give qualitatively the   same results (i.e.\  the emergence  of jam
waves).   For  every   (arbitrary) configuration of    the  model, one
iteration  consists of the following steps,  which  are each performed
simultaneously for all vehicles:\bi

\y {\bf Acceleration of free vehicles:} Each vehicle of speed $v <
v_{max}$ whose predecessor is $v+2$ or more sites ahead, accelerates
to $v+1$: $v \to v + 1$.

\y {\bf Slowing down due to other cars:} Each vehicle (speed $v$)
whose predecessor is $d=v$ or less sites ahead, reduces its speed to
$d-1$: $v \to d-1$.

\y {\bf Randomization:} Each vehicle (speed $v$) reduces its speed
by one with probability $1/2$: $v \to \max[ \, v-1 , 0 \,]$ (takes
into consideration individual fluctuations).

\y {\bf Movement:} Each vehicle advances $v$ sites.

\ei
The three first steps may be called the ``velocity update''.  They
have been constructed in a way that no ``accidents'' can happen during
the vehicle motion.

A comparison with real traffic measurements~\cite{NaS92}
indicates that it is reasonable to
assume that, at least to the order of magnitude, one site occupies
about $7.5~m$ (which is the space one car occupies in a jam), one
iteration is equivalent to about $1$~second, and maximum velocity $5$
corresponds to about $120$ km/h.

\sec Density waves

In a closed system (periodic boundary conditions, i.e.\ ``traffic in a
closed loop''), the number~$N$ of cars and therefore the density
$\bar\rho = N/L$ are conserved ($L$: system size).    Average quantities
such as throughput $q$ are then functions of $\bar\rho$.  Our model
reaches its maximum throughput $q_{max} = 0.318 \pm 0.001$ at a
density of $\rho^* = 0.086 \pm 0.002$ (Fig.~\ref{fdiag}{1}).

But what  is the  deeper  reason behind  this capacity  threshold?  In
order  to  access   this  question,  as  a   first  step  we  look  at
space-time-plots of   systems slightly below   and above the threshold
density $\rho^*$ (first row  of Fig.~\ref{resolution}{2}).  Similar to
the usual plots of 1d cellular  automata, in these pictures horizontal
lines are   configurations  at consecutive time   steps, time evolving
downwards.    Black pixels stand for   occupied  sites.  Vehicles  are
moving from left  to  right, and by    following the pixels, one   can
discern the trajectories of the vehicles.

These pictures show marked shock waves, and  they occur more often for
the higher density.  These waves form at arbitrary times and positions
due to a ``bad''   superposition of the   disturbances caused by   the
random noise of  the velocity  update.  They  are clearly visible   as
clusters of cars of low velocity  (with more interior structure inside
the clusters). Once such a disturbance has formed, it is maintained as
long as there are more vehicles arriving at the end  of the queue than
vehicles leaving the queue at its head. These disturbances appear well
{\it below\/} the range of maximum traffic capacity, but they are rare
and  only start to dominate  the system's appearance  at densities far
above the range of maximum capacity.  This  leads to the idea that the
range of maximum traffic flow might be reached when there are, for the
first time,  waves  with  a  ``very  long''  lifetime, similar  to   a
percolation   transition~\cite{Sta92}.   (See  also~\cite{TT}    for a
similar analysis of a deterministic model.)

To get a better overview, the second and third row of
Fig.~\ref{resolution}{2} show the same system at lower resolutions
obtained by averaging, therefore showing a larger part of the system
and more time steps.  A striking feature of these pictures is that they
look in some way self-similar~\cite{Mandelbrot,Feder},
i.e., large jams are composed of many
smaller ones which look like large ones at a higher resolution.

A two-lane model
gives similar results~\cite{NaR93}.
We therefore assume that results from the single-lane model can
be taken over to the more realistic case.

\sec Self-organization of maximum throughput

A second reason for looking especially at the threshold density
$\rho^*$ is that it self-organizes as the outflow from a jam.  In
order to see this, in a system of length~$L=10^6$, we filled the left
half with density~$\rho_{left}=1$ and left the right half of the
system empty: $\rho_{right}=0$ (cf.~Fig.~3).  We used an open boundary
condition at the right, i.e., vehicles on sites $L-v_{max}, \ldots, L$
were deleted.  The left boundary was closed.

We then ran the system according
to the update rules.
%~(see Fig.~\ref{outflow.pic}{6}).
After
$t_0 = 2 \cdot 10^5$~time
steps we started to count the vehicles which left the system at the
right boundary.  In Fig.~\ref{outflow.val}{4} we show the average
throughput
\[
q_{open} = {1 \over t - t_0} \, \int_{t_0}^t \d t' \, n \ ,
\]
which is $0.318 \pm 0.01$ for large times.  This is, within
errors, exactly the value of maximum throughput $q_{max}$ for the
closed system.  In addition, even when filling up the left half
of the system randomly only with a much smaller
density $\rho_{left} = 0.1$,
the outflow is the same.
We conjecture therefore that the {\it outflow\/}
from a high density regime selects by itself the state of
maximum average
throughput; and ``high density'' means an average
density above the threshold density $\rho^*$.

This is comparable to the case of boundary-induced state
selection for asymmetric exclusion models~\cite{Krug}, with one
significant difference: Our model does {\it not\/} select this
state of maximum throughput when adding as many particles as
possible at the left boundary~\cite{NaS92}---the
particles have, at the start of the simulation,
already to be inside the system as described above.  We show
in~\cite{NaH93} for a simpler model that this only can be overcome by
some artificial update rules for a few sites at the left
boundary; the same is true for the model here.

\sec Lifetimes and cluster labeling

We have indeed  measured fractal properties of the  plots in  order to
test for self-similarity, but in  the following we  want to present  a
quantity whose results we have found easier to interpret: the lifetime
distribution of the jams. Lifetime is the number of time steps between
the first and the last time some car has to slow  down due to the same
disturbance.

After    the deterministic part   of    the  update  and before    the
randomization step, all ``free'' cars have velocity $v = v_{max}$.  We
therefore   define all cars   with  $v <   v_{max}$  at this point  as
``slow''.  We then looked for  ``clusters'' of slow  cars in the model
and  measured the  lifetime of  these   clusters.  The idea itself  is
borrowed from avalanche models~\cite{Bak}, but in the traffic model it
is not  possible to wait until  one avalanche ($=$ jam)  is dissipated
before  originating the next one.  We  therefore had to  keep track of
multiple traffic jams in the model simultaneously.

Technically, we distinguished different jams by different labels,
and the jam of each label
$lbl$ was active between $t_{start}(lbl)$ and $t_{end}(lbl)$.
Initially, we set $t_{end} = 0$ and $t_{start} = t_{max}$ for all $lbl$.
($t_{max}$ is the total number of iterations of the simulation run.)
Then, at each time step after the deterministic and before the random
part of the velocity update, we did the did the following: \bi

\y All ``fast'' cars get a very high label number $lbl_{max}$, with
$t_{start}(lbl_{max}) = 0$ and $t_{end}(lbl_{max}) = t_{max}$.

\y Being ``slow'' (in the sense of the above definition)
can in the model only be caused by two reasons:
Either the car $n$ had to slow down because the next one ahead $n+1$
was too close, or the car has not yet accelerated to full speed due to
a jam which it just has left.  Therefore,
\[
t_{start}(n,t) = \min[ t_{start}(n+1,t-1) , t_{start}(n,t-1) , t ] \ .
\]
In words, this means that if two different jams may be the origin of
$n$'s slowness, then the algorithm selects the older one.

\y Then the label is set to the one of the selected jam:
\[
\vtop{\halign{&#\hfil\cr
IF $t_{start}(n,t)=t_{start}(n+1,t-1)$ THEN \cr
\qquad $lbl(n,t) = lbl(n+1,t-1)$ \cr
ELSE IF $t_{start}(n,t)=t_{start}(n,t-1)$ THEN \cr
\qquad $lbl(n,t)=lbl(n,t-1)$ \cr
ELSE \cr
\qquad $lbl(n,t) = newlbl$ \cr
ENDIF \cr
}}
\]
where $newlbl$ is a new label not yet used.

\y
Next, $t_{end}$ is updated: $t_{end}(lbl(n,t)) = t$.

\ei
The overall result of this labeling is that every
vehicle which becomes ``slow'' without another ``slow'' car
as a cause originates a new jam with an associated lifetime.
When one jam splits up into several branches, they all
obtain the same label because they have the same origin.
In consequence, only the branch
which stays ``alive'' the longest time determines the lifetime
of this specific jam.
When two branches {\it completely\/}
merge together, the ``older'' one
takes over.  The younger one then no longer exists, but it is
counted for the statistics because it had its own independent
origin.

We implemented this algorithm on a Parsytec GCel-3 parallel
computer, where we could use up to 1024~processors.  The
dynamics itself was implemented in a
``vehicle-oriented'' way which means
that we had a list of positions $(x_i)_i$ and a list of
velocities~$(v_i)_i$ for the vehicles~$i=1,\ldots,N$.  As
passing is not allowed in the single lane model, this list
always remains ordered.  We therefore could
distribute the model by
placing $N/p$ consecutive vehicles
on each of the $p$ processors.  This resulted, for large system
sizes, in a computational
speed of $8.5 \cdot 10^6$ {\it particle\/}-updates per second on
512~processors, compared to $0.34 \cdot 10^6$ on a Sparc10
workstation.  (At a density of $\rho = 0.08$, this corresponds
to $106 \cdot 10^6$ resp.\ $4.25 \cdot 10^6$ {\it
site\/}-updates per second.)  But for a smaller system size of
$L=10^5$ ($\rho=0.08$)
the computational speed on 512~nodes decreased to $3.1
\cdot 10^6$ particle-updates ($= 39 \cdot 10^6$ site-updates) per second.

For  the parallel cluster  labeling,  our implementation followed  the
idea of~\cite{FlT93}.  That means that labels were assigned locally on
the processors, and only labels that touched boundaries were exchanged
with  the neighbors.  After  the  labeling, information on  ``active''
labels was exchanged by a  relaxation method (see~\cite{FlT93}) to the
leftmost processor which has  this label in use.    By this method  we
kept  track of ``active'' jams, and  lifetimes of ``dead'' jams (i.e.\
cluster labels which were no longer in use) could be recorded.

For sufficiently large system sizes, the computational speed
went down by a factor of four due to the labeling; but for smaller
systems of size $L = 10^5$ and $\rho \approx 0.08$, 512
processors were inefficient.  The following table shows
computational speeds (in MUPS $=$ MegaUpdates Per Second
$= 10^6$~site-updates per second)
for these parameters ($\rho=0.08$):
\[\vtop{\offinterlineskip
\halign{&#\hfil&\quad\strut\vrule#\quad\cr
number of computational nodes
&& 32	&& 128	&& 512 \cr
&& ~ 	&&	&& \cr
\noalign{\hrule}
&& ~	&&	&& \cr
speed w/o labeling
&& 6.8	&& 27	&& 106 \cr
speed with labeling
&& 1.5	&& 5.5 	&& 5.7 \cr
}}
\]
In consequence, we usually used 128 processors per job.
About five days of computing time on 512~processors ($4 \times 128$)
were needed for
the results presented in this paper.

\sec Results of lifetime measurements

Fig.~\ref{0.5}{5} (lower branch)  shows  the results for the  lifetime
distribution of our traffic  model.  The figure shows the (normalized)
number  $n$ of   traffic jams of   lifetime  $\tau$;  as  the  data is
collected   in   ``logarithmic  bins'',   the  $y$-axis   is therefore
proportional to $\tau \cdot n$.  For a  density of $\rho = 0.08$ (near
the capacity   threshold density $\rho^*$), there  is   a region where
$n(\tau)   \propto \tau^{-\alpha_1}$ ($\alpha_1 =   3.1  \pm 0.3$) for
$\tau$ approximately between $5$ and $50$, and another region where $n
\propto
\tau^{-\alpha_2}$    ($\alpha_2  =     1.65 \pm     0.08$)  for $\tau$
approximately  between $100$  and $5000$. For a  higher density of  $\rho =
0.1$, the second  regime  gets  slightly  longer whereas it   vanishes
totally  for a lower density  of $\rho =   0.06$.  In other words, the
change-over  from the  light   traffic regime ($\rho   < \rho^* \equiv
\rho(q_{max})$) to  the heavy  traffic   regime ($\rho >  \rho^*$)  is
accompanied by a   qualitative  change in  the   lifetime distribution
(i.e.,  the emergence of a  regime with $n \sim \tau^{\alpha_2}$), but
the lifetime distribution does not show critical behavior in the sense
of a percolation transition because of the upper cut-off.

This cut-off of the lifetime distribution near $\tau =
500000$ is not a finite size effect.  Since we analyze clusters in
a space-time-domain, finite size effects could be caused by
space or by time.
For the space direction, we have in Fig.~\ref{0.5}{5}
superimposed the results for system sizes $L = 10^4$ and
$L=10^5$.  The scaling region is not any longer for the larger
system.  For the time direction, we have measured the
third moment
$
\mede{\tau^3} := \left. \int \d \tau \, \tau^3 \, n(\tau)
\right/ \int \d \tau \, n(\tau)
$
of the lifetime distribution as a
function of time.  For a critical (or supercritical)
distribution, this moment should diverge with time.
And indeed, we find that approximately $\mede{\tau^3} \propto
t^2$ for sufficiently small $t$.
But for large enough $t$
($\approx 10^4$ for $\rho = 0.08$), $\mede{ \tau^3 }$ becomes
constant and therefore independent of $t$.  This means that
longer simulation times would not lead to (on average)
longer lifetimes.  In consequence, the cut-off in the lifetime
distribution is no finite time effect.

In order to find out if these results depend on our cluster
labeling technique, we also implemented for comparison
a Hoshen-Kopelman cluster
labeling~\cite{Hoshen-K}, which labels different jams as being the same
already when they only ``touch'' each other.  This should therefore
lead
to longer lifetimes.  Nevertheless, this leads qualitatively
to the same results;
but as we did not implement this second method on
the parallel computer, data quality was not high enough
to make quantitative comparisons.

\sec Reducing the fluctuations at maximum speed

An intuitive explanation of our findings of the last section
might be as follows:  Random superposition of velocity fluctuations
leads to the formation of a wave.  Once this wave has formed, it
is relatively stable and may therefore seen as a
collective phenomenon.  Indeed, experiments without any noise~\cite{NaH93}
show that a wave lasts forever when the density is above
$\rho^*$.  But with noise, another wave
may form further upstream, and the outflow from this wave may be
low enough over a certain period of time so that the original wave
dissolves.  By this mechanism, the criticality of the
deterministic model is destroyed by the noise.

If this argument were true, then a reduction of only the fluctuations
at high speed should extend the scaling regime.  For this
purpose, the ``randomization'' step of the update algorithm was
replaced by the following rule:\bi

\y {\bf New randomization:} If a vehicle has maximum speed
$v_{max}$, then it reduces its speed by one with a much lower
probability $p_{fluc} = 0.005$.
Otherwise, it reduces its speed by one with
probability 0.5 (as before).

\ei
By this rule, only the fluctuations at $v = v_{max}$ are changed,
whereas the slowing down or the acceleration remain the same.

The part of the fundamental diagram (throughput
versus density)
near the throughput maximum
is included in Fig.~\ref{fdiag}{1}.  The maximum
throughput becomes slightly higher
for this new model
and is found at a somewhat
lower density, but the change in throughput is only $2\%$.

In the scaling plot of the lifetime
distribution (Fig.~\ref{0.005}{5}), the scaling region of the
``second'' regime clearly gets longer and extends now over about
three orders of magnitude from $\tau=200$ to $\tau=200000$.
In this region, $n \sim \tau^{-\alpha'_2}$ with $\alpha'_2
= 0.55 \pm 0.05$, which is different from the value before, but still
within error bars.

Our interpretation of this is that jams are indeed more rarely
dried out by other jams forming upstream, which makes longer
lifetimes possible.  Or in other words: The typical
length scale $\lambda$ between jams becomes larger with smaller
$p_{fluc}$ and should diverge for $p_{fluc} \to 0$.
Plots of the space-time-domain (not shown
here) confirm this interpretation.

The limit $p_{fluc} \to 0$ is
singular: For $p_{fluc} = 0$ and $\rho < \rho_c(p_{fluc}\!=\!0)
\equiv 1/(v_{max}+1)$ the
closed system will eventually settle down in a state where all
vehicles move with velocity $v_{max}$.  Then the noise in the
acceleration or slowing down does no longer play a role, and the
model reduces to the light traffic regime of~\cite{NaH93}.  The
maximum average throughput then is
\[
q_{max}(p_{fluc}\!=\!0)
= \rho_c \cdot v_{max} = { v_{max} \over v_{max}+1 } \ ,
\]
i.e.\ $q_{max} = 5/6  \approx 0.833$ for $v_{max} =  5$, which is more
than twice the  values for $p_{fluc}  > 0$.  And if  one  takes $L \to
\infty$ before taking $p_{fluc} \to 0$,  then the point $p_{fluc} = 0$
cannot be approached continuously: Some fluctuation will always create
a jam which redistributes the vehicles at a lower density.  These
observations are consistent with the bistable state for $\rho$ between
$1/3$ and $1/2$ in the model of Takayasu and Takayasu~\cite{TT}.

This  means that for $p_{fluc}  \to 0$ the   system separates into two
phases, into regions of laminar  traffic where $v \equiv v_{max}$ and
$\rho <  \overline{\rho}$,  and into   dense regions  with many   jams
(``turbulent'') where $v   << v_{max}$ and  $\rho >> \overline{\rho}$.
The density   in the laminar regime is   totally determined by the
outflow  from the  jam and much   lower than  what  could be  reached if
$p_{fluc} \equiv 0$.

These findings are confirmed by a short calculation.
If one considers the outflow of the vehicles out of a
dense jam, one notices that the first vehicle starts at
time step~1 with probability $P(t=1) = 0.5$ ($0.5$ is the
value for the fluctuations during acceleration),
at time step~2 with
probability $P(t=2) = (1 - P(t=1)) \cdot 0.5 = 0.5^2$ and therefore
at time step~n with probability $P(n) = 0.5^n$.
In consequence, the average time between two vehicles being
released is
\[
\overline{t} = \sum_{n=1}^\infty n \cdot 0.5^n = 2 \ .
\]
Since, in addition, the front of the jam moves backwards, the
throughput at a fixed point is {\it at most\/}
$1 / \overline{t} = 1/2$,
which is much lower than the maximum throughput of $q = 5/6$
for $p_{fluc} \equiv 0$.

\sec Possible consequences for real traffic

When discussing the relevance for traffic, one should bear
in mind that real traffic with its mixture of different vehicles,
different road conditions, and on more than one lane
certainly is much more complicated
than the model discussed here.  Nevertheless, the model should
not only give realistic answers for the single-lane traffic
considered here, but should as well be valid for homogeneous
multi-lane traffic streams (as arising in, e.g., commuter
traffic).

The first observation from our model then is that the stability
or instability of the ``fast'' regions do not change the
overall flux considerably.  In other words: For the
throughput it does not
matter if there is one jam-wave with a very long lifetime, or a
multitude of short-lived ones.  Therefore, at least for
homogeneous traffic situations (vehicles not too different),
keeping the speed constant (as by cruise control) does not
enhance the throughput, although it certainly leads to a more
convenient driving because jams become rare.

In addition, our findings may be seen in the light of the
``capacity drop issue''~\cite{PH88}.
This means the observational fact that
streets can, for short times, support much higher
traffic loads than for longer time averages.  In the framework
of our model the interpretation is as follows:  If traffic does not
flow out of a jam but aggregates by some other mechanism,
relatively high loads are possible until a fluctuation due to
$p_{fluc}$ leads to a jam which redistributes the vehicles at a
lower density and with lower throughput.

The relevant question for traffic engineering in this context
is how to define the capacity ($=$ maximum throughput) of a street.
In the sense of statistical physics it is obvious that for $L
\to \infty$ only the lower value of $q(p_{fluc}\!>\!0)$
is relevant; but what is the truth for reality  with its finite length
and time  scales?  Our  findings indicate that   one should extend the
length scale $\lambda$ which  is presumably  associated with $p_{fluc}$
beyond the length of some critical parts of the road, e.g.\ the length
$L_{bn}$  of a bottleneck  due to construction  work.  The result then
would be that there may be days when no jam forms although the load is
higher than $q_{max}(L\!\to\!\infty)$, whereas when $\lambda < L_{bn}$
one would have a  jam nearly every day.  On  the other hand, if it  is
{\it not\/} possible  to extend  $\lambda$ far  enough, then,  for the
situation of homogeneous traffic described  here, technical measures as
described here are  most  probably irrelevant for the  throughput, and
the decision is only one of safety and convenience.  See
Ref.~\cite{TRB} for a broader survey in how far specific alterations
of the driving behavior can change throughput.

If differences between vehicles become relevant (e.g.\ for a
mixture of trucks and passenger cars and without speed limit),
then the situation becomes more complicated. Results for
multi lane traffic,
together with different types of vehicles,
are the subject of a forthcoming publication~\cite{NaR93}.

\sec Summary

We have investigated traffic jams which emerge in a natural way
from a rule-based, cellular-automata-like traffic model
when operating the model near the maximum traffic throughput.
The model includes strong driving by noise, taking into account
the strong fluctuations of traffic.  In
space-time-plots, the jams have a self-similar appearance, and
the numerically determined
lifetime distribution indeed shows a scaling region over
about one and a half orders of magnitude for
systems near the maximum throughput.  But the cut-off
towards longer lifetimes was shown to be {\it no\/} finite size
effect, so that the lifetime distribution does not indicate
critical behavior in its strict meaning.

However, when reducing only the fluctuations $p_{fluc}$ at maximum speed
(and not those of slowing down or acceleration)
by a factor of 100, the scaling
region was shown to become longer and now to extend over
three orders of magnitude.  It was argued that the
limit $L \to \infty$, $p_{fluc} \to 0$ is singular in the sense that it is
different from the limit $p_{fluc} = 0$, $L \to \infty$.

In addition, when the driving by boundary conditions is
strong enough, the system selects automatically
the density of its maximum throughput, which is identical to the density
where the scaling first appeared.
In the limit of $p_{fluc} \to 0$,
this may be seen as an example of self-organized
criticality~\cite{Bak}.

Implications for traffic include that technical measures to
reduce fluctuations such as
cruise control only have an effect on maximum throughput when
the fluctuations can be suppressed under a certain level related
to the length of a bottleneck; otherwise, improvements will
only be in safety and convenience.

\sec Acknowledgements

I want to thank A.~Bachem, H.~J.~Herrmann, T.~Poeschl,
M.~Schreckenberg, D.~Stauffer, D.E.~Wolf, and A.~Schleicher, who in
addition contributed the data for Fig.~4.  I acknowledge computing
time on the Parsytec GCel-3 of the Zentrum f\"ur Paralleles Rechnen of
the University of Cologne.  The referee made helpful suggestions.  I
finished this work in the TRANSIMS group at Los Alamos National
Laboratory and at the Santa Fe Institute, and I thank them for
hospitality.

\vfill\eject\section*{Figure captions}
\vskip-\bigskipamount
\def\fig#1#2{\bigskip\noindent{\bf Figure #2:\hfill\break}}

\fig{fdiag}{1}
Parts  of   the  fundamental  diagrams  (i.e.\ throughput~$q$   versus
density~$\rho$) near the capacity maximum for the original model ($+$)
and for the model with reduced  fluctuations (squares) presented later
in this text.

\fig{resolution}{2}
Plots    of  simulated  single   lane    freeway   traffic  in     the
space-time-domain with resolutions   (a)\  1:1, (b)\ 1:4,   (c)\ 1:16.
Vehicle density   $\overline{\rho}   =   0.07$   (left    column)  and
$\overline{\rho}  = 0.1$ (right).  Each black  pixel  corresponds to a
site occupied by a vehicle at a certain place ($x$-direction) and at a
certain time ($y$-direction).   A trajectory of an undisturbed vehicle
goes therefore diagonally downwards and to the right.  The pictures of
the first row cover $500$~sites and $500$~time steps.

The pictures of each row are contained (as indicated by the boxes)
in the pictures of the row underneath.

\fig{outflow.pic}{3}
Space-time plot of the outflow from a jam (see text).  As in Fig.~2,
the horizontal direction is the space direction and time is running
downwards.  The system size is $L=500$, much smaller than the systems
used for Fig.~4.

\fig{outflow.val}{4}
Average outflow (see text) from a high density region as a function of
time. The  straight  line shows the  outflow with  $\rho_{left}(t=0) =
1.0$,   the   broken line  shows    the outflow  from a   region  with
$\rho_{left}(t=0)  = 0.1$.  In  both  cases, the system self-organizes
towards the state of maximum throughput.

\fig{0.5}{5}
Comparison of  lifetime distributions~$n(\tau)$  for the  traffic jams
between the ``standard'' model and the ``model with cruise
control'' (upper branch).
The data is collected in logarithmic bins, therefore the $y$-axis is
proportional to $\tau \cdot n(\tau)$, and it has been normalized such
that $n(\tau=1) = 1$ for the lower and $n(\tau=1)=100$ for the upper
branch.
{\it Lower branch\/}:   standard  model, i.e.\  $p_{fluc}  = 0.5$.
Straight lines, from  left to right:  Results for system size $L=10^5$
and densities $\rho =  0.06$, $0.08$, and  $0.10$, i.e.,  below, near,
and  above the threshold  density~$\rho^*$.  Dotted lines: Results for
same densities, but smaller system size $L=10^4$.
{\it Upper branch\/}: including  ``cruise control'', i.e.\ $p_{fluc} =
0.005$.  System size $L=10^5$, densities $\rho = 0.09$, $0.08$,
$0.07$, and $0.06$, as noted in the legend.


\begin{thebibliography}{999}
\bibitem{PH88}
Persaud B N, Hurdle V F, \tit Some new data that challenge some old ideas
about speed-flow relationship. \jrn{Transp.\ Res.\ Rec.} \rf 1194 88
191--198.
\end{thebibliography}
\end{document}